\documentstyle[11pt,appb,epsfig]{article}  

\newcommand{\beq}{\begin{equation}}
\newcommand{\eeq}{\end{equation}}
\newcommand{\bea}{\begin{eqnarray}}
\newcommand{\eea}{\end{eqnarray}}

\newcommand{\gsim}{\stackrel{>}{\sim}}

\begin{document}
\title{Medium Dependence of the Vector-Meson Mass: Dynamical and/or 
Brown-Rho Scaling? \thanks{Presented at the International Workshop
on 'The Structure of Mesons, Baryons, and Nuclei' on the occasion
of Josef Speth's 60th birthday; to be published in Acta Physica Polonica
{\bf B}} }
\author{G.E. Brown, G.Q. Li, R. Rapp, \address{Department 
of Physics and Astronomy, \\
State University of New York at Stony Brook, \\
Stony Brook, NY 11794-3800, U.S.A.\\}
\hspace{1cm} \\
Mannque Rho, \address{Service de Physique Th$\acute {\rm e}$orique, 
CEA Saclay,\\ F-91191 Gif-sur-Yvette, France\\}
\hspace{1cm} \\ 
and J. Wambach \address{Institut f\"ur Kernphysik,
Technische Universit\"at Darmstadt,\\
Schlo{\ss}gartenstr. 9,
D-64289 Darmstadt, Germany}  
}
\maketitle
\begin{abstract}
We discuss the similarities and differences for the 
theories of Rapp, Wambach and collaborators (called R/W in short)
and those based on 
Brown-Rho scaling (called B/R), as applied to reproduce the 
dileptons measured by 
the CERES collaboration in the CERN experiments. In both theories 
the large number of dileptons at invariant masses $\sim$~$m_\rho/2$ 
are shown to be chiefly produced by a density-dependent $\rho$-meson
mass. In R/W the medium dependence is dynamically calculated 
using hadronic variables defined in the matter-free vacuum. In B/R scaling 
it follows from movement 
towards chiral symmetry restoration due to medium-induced 
vacuum change, and is described in terms
of constituent (or quasiparticle) quarks. 
We argue that the R/W description should be reliable up to densities 
somewhat beyond nuclear density, where hadrons are the effective 
variables. At higher density there should be a crossover to 
constituent quarks as effective variables
scaling according to B/R. In the crossover region, the two descriptions
must be ``dual.''\\
For the moment there is a factor $\gsim$~2 difference between the 
predicted number of dileptons from the two theories, B/R scaling 
giving the larger number. We show that a substantial factor results
because in B/R, fluctuation is made about the ``vacuum'' modified by
density, so that a different
mass $m_\rho^*$ appears in the Lagrangian for each density, thereby
rendering residual
interactions between hadrons weaker, whereas R/W 
calculate a mass $m_\rho^*$ for each density with an effective Lagrangian
defined in the zero-density vacuum, which has the free $m_\rho$ in the 
Lagrangian and hence the coupling is strong. Thus more diagrams need to be
incorporated in R/W to reduce the discrepancy. On the other hand, 
R/W include processes which may be additional to these of B/R. These
constitute several (smaller) corrections. \\ 
It is argued  that the $N^*$-hole
state $[N^*(1520) N^{-1}]^{1^-}$ is almost completely $\rho$-meson 
like in content; {\it i.e.}, it is, to a good approximation, just the 
state $\rho|0\rangle$ that would be produced by the $\rho$-meson
field acting on the nuclear ground state (finite temperatures are
not expected to disturb this picture by much).   
\end{abstract}
  
\section{Introduction}
Relativistic heavy-ion collisions carried out at the CERN-SpS by the 
CERES collaboration~\cite{sau} show an excess of dileptons in the mass region 
below the vector meson mass $m_\rho$. For dilepton invariant masses of
$\sim$~$m_\rho/2$, the enhancement is an order-of-magnitude over 
conventional theories which do not have a medium-dependent $\rho$-meson 
mass. In fact, roughly speaking, the $\rho$-meson must have an in-medium 
mass of $\sim$~$m_\rho/2$ for a sufficient time in 
the heavy-ion collision to be able to
produce the enhanced number of dileptons of invariant mass 
$\sim$~$m_\rho/2$. Such times are provided in current transport
calculations of the CERN experiments.  

We argue in the following that this can be accomplished in the meson
sector by a dynamical lowering of the $\rho$-meson mass produced by
coupling to $N^*$-hole states, especially the 
$[N^*(1520) N^{-1}]^{1^-}$. This is an important ingredient in the 
R/W theory~\cite{RCW} which makes it possible to approximately describe 
the number of dileptons in the mass region $\sim$~$m_\rho/2$.
The R/W theory posits hadrons with free 
masses as the appropriate variables. This must be reliable at
low density. But at high density, the R/W theory will
become strong-coupling necessitating the consideration of a large number of
diagrams. It is then more advantageous to go over to a 
medium-modified ``vacuum'' around which fluctuations into various
flavor directions can be treated in a weak-coupling approximation.
This is the basis of the quasiparticle picture associated 
with B/R scaling~\cite{BRscaling,LKB}, suggested also by QCD sum-rule 
calculations~\cite{HaLe,JiLe}. 

Although we use the shorthand notation for this state 
$[N^*(1520) N^{-1}]^{1^-}$, it is a highly collective state, in which
each of the $N$ nucleons is excited, with coefficient $1/\sqrt{N}$, 
to the $N^*$ at energy 1520~MeV.
Since the 580~MeV excitation energy ([1520-940]~MeV) is large compared 
with the temperatures encountered in heavy-ion collisions, the collective 
state is relatively unaffected by temperature. In 
the hot fireball many excitations of nucleons such as the $\Delta(1232)$
isobar will be present. We argue that, in analogy with nuclear structure,
these excited states of the nucleons should have dipole excitations at
not very different relative energies from that of the dipole excitation 
built on the ground state. By retaining only
nucleons, as Rapp and Wambach have done, we underestimate the dynamical 
change in the $\rho$-meson mass. This should be kept in mind when considering
central heavy-ion collisions where their present theory seems to 
underestimate the experimental data. Inclusions of the dipole excitations 
built on nucleon excited states would substantially increase their
predicted dilepton yield. An upper limit to this effect (R/W
now have a lower limit) would be obtained by using the total baryon
density rather than that of just the nucleons, as they have done.   

Brown, Buballa and Rho~\cite{BBR} showed that the chiral phase 
transition could be sensibly made only if the relevant variables close
to the transition were constituent quarks, the chiral restoration being
accomplished in terms of the quarks going massless in the chiral limit.
On the other hand, nuclear matter is more economically described in 
terms of hadron variables to 
describe saturation although hybrid descriptions in terms of hadron-quark-bags
and hadrons with B/R scaling can be constructed to work~\cite{SBMR}. 
Somewhere between 
nuclear matter density and that 
of chiral restoration, the effective variables will smoothly 
change from hadronic to constituent quarks subject to B/R scaling. 
(This change over
was not, however, constructed by Brown, Buballa and Rho.) 
This construction of the phase transition suggests that the 
Rapp/Wambach description should hold for 
densities up to and somewhat beyond
nuclear matter densities, with hadrons as the relevant variables,
but give way to that from Brown/Rho scaling at the higher densities. 
Near the hadron-quark changeover, the 
two descriptions must be ``dual'' to each other in a way analogous
to quark-hadron duality in heavy-light-meson systems~\cite{qhduality}. 
In this description, the dynamical lowering obtained by R/W
may give way at higher density 
to the B/R scaling dictated by chiral (and scale) symmetry 
before reaching the chiral restoration transition. 

The dilepton production, which is the only way we presently 
have of mapping out the $m_\rho^*$ as function of density, 
should be insensitive to the density at which this crossover is made, 
and we suggest how this can be realized in practice. 
 
In this note we deal only with the density dependence in the
$\rho$ meson-mass. However, symmetries in (constituent) quasiquarks
are expected to give a similar scaling for the $\omega$ as well as
other light-quark hadrons.
Work on this will be deferred to a future investigation.

It should be mentioned that there is evidence for this ``dual'' description
at density up to that of normal nuclear matter. It was shown in 
\cite{FR96,SBMR} that the B/R scaling parameter (denoted as $\Phi$) is related
by Landau Fermi-liquid theory to the Landau quasiparticle interaction 
constant $F_1$.
This connection is a statement on a possible relation at
low energy/density between the vacuum structure characterized by
the quark condensate and the nuclear interactions characterized by
Landau Fermi-liquid parameters. The upshot of the present paper is that  
beyond nuclear matter density up to the chiral phase transition, 
Fermi liquid theory of quasiquarks may become 
more appropriate. Such phenomena as
color superconductivity and flavor-color locking discussed 
recently~\cite{wilczek} could be treated more realistically starting from the
Fermi liquid theory of quasiquarks.    
 
\section{Dynamical mass scaling} 

Most interesting for the dynamical effects are the invariant mass 
distributions at small and large $p_\perp$ of electron pairs.  
Since the CERES acceptance is basically  central rapidity, 
$p_\perp \simeq p$, the total momentum, so that the energy $p_0$ 
of an electron pair is essentially
\beq
p_0\simeq \sqrt{(m_\rho^*)^2+p_\perp^2} 
\eeq 
with $m_\rho^*$ the effective mass of the vector mesons $\rho$ or $\omega$. 
Experimentally it turned out that the low $p_\perp$ region is highly enhanced, 
by a factor $\sim$~10, whereas the high $p_\perp$ region seems to exhibit 
a much  smaller enhancement~\cite{Voigt}. This is completely 
opposite to fig.~7 of
Friman and Pirner~\cite{FrPi}. Since the Friman-Pirner paper it has
been realized that the cutoff in the form factor used by these authors
($\Lambda_{\rho N^*N}$=1.5~GeV) is too large. Thus the $P$-wave coupling 
to $N^*$ resonances was overestimated. Similar conclusions were reached by
Friman for the $\pi NN$ form factor~\cite{Fr97} (which is related to  
medium effects in the pion cloud of the $\rho$), which can be severely 
constrained from the analysis of $\pi N\to \rho N$ data. 

An important paper by S.H. Lee~\cite{Lee} shows that from QCD sum
rule calculations the vector meson mass, broken down into scalar
and momentum-dependent components, is for the $\rho$-meson 
\beq
\frac{m_\rho(n_n)}{m_\rho(0)} =1-(0.16\pm 0.06) \frac{n_n}{n_0} 
-(0.014\pm 0.005) \left(\frac{\vec{q}}{0.5}\right)^2 \frac{n_n}{n_0} \ , 
\label{scaling}
\eeq
where $|\vec{q}|$ is in GeV and $n_0$ is the nuclear saturation density. 
Thus, the scalar term, which he interprets as Brown/Rho scaling, 
is large, whereas the ${\vec q}^2$ term, which could arise from $P$-wave
coupling of the $\rho$ to isobars and from relativistic corrections
to $S$-wave couplings, is small. This is in agreement
with the $p_\perp$-distribution. 
Rapp's calculation for $p_\perp < 0.4$~GeV/c, compared with the
experimental points is shown in the left panel of fig.~1 (Dalitz
pairs are not included here).
\begin{figure}
{\makebox{\epsfig{file=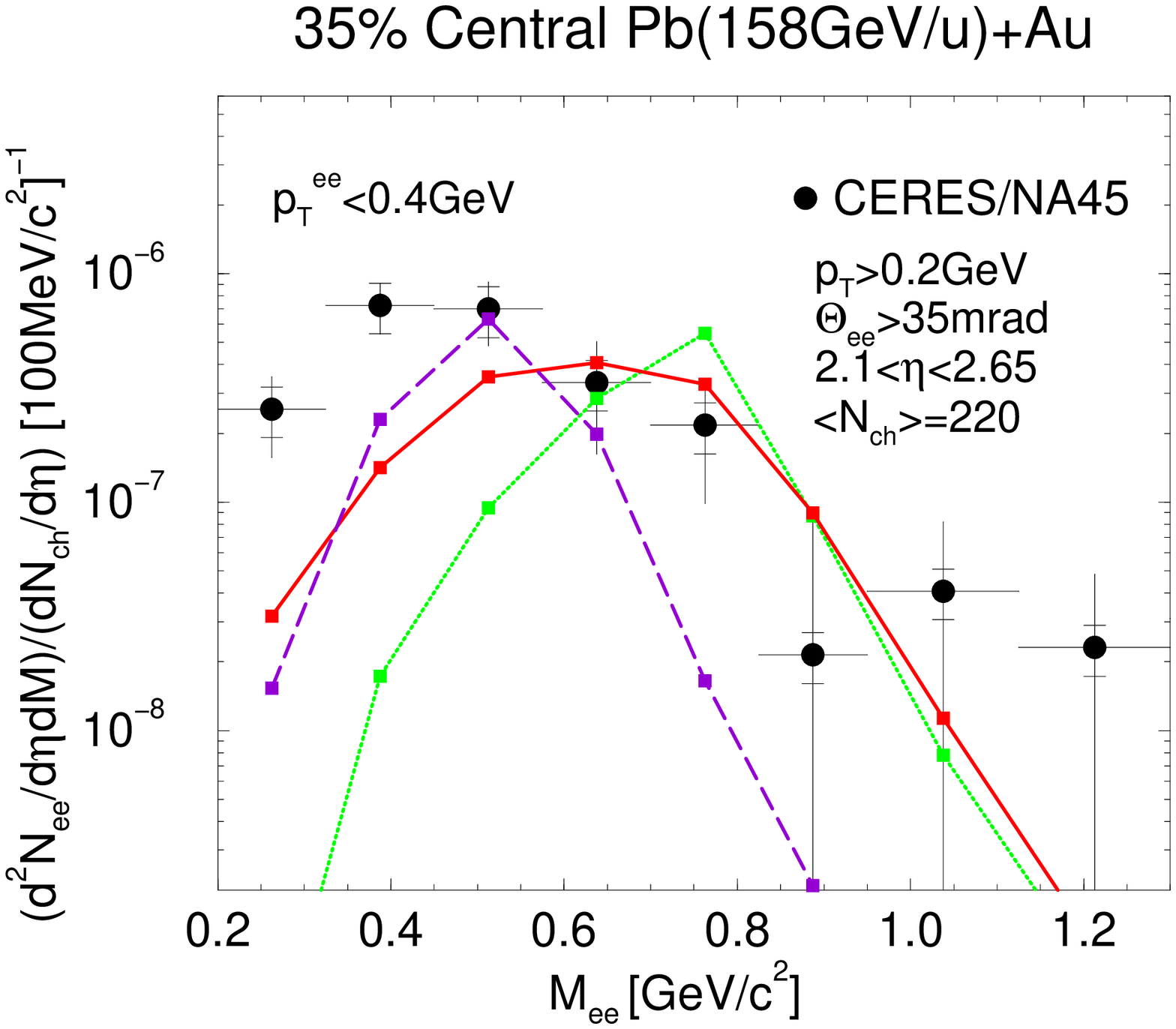,width=5.7cm,height=6cm,
}}}
{\makebox{\epsfig{file=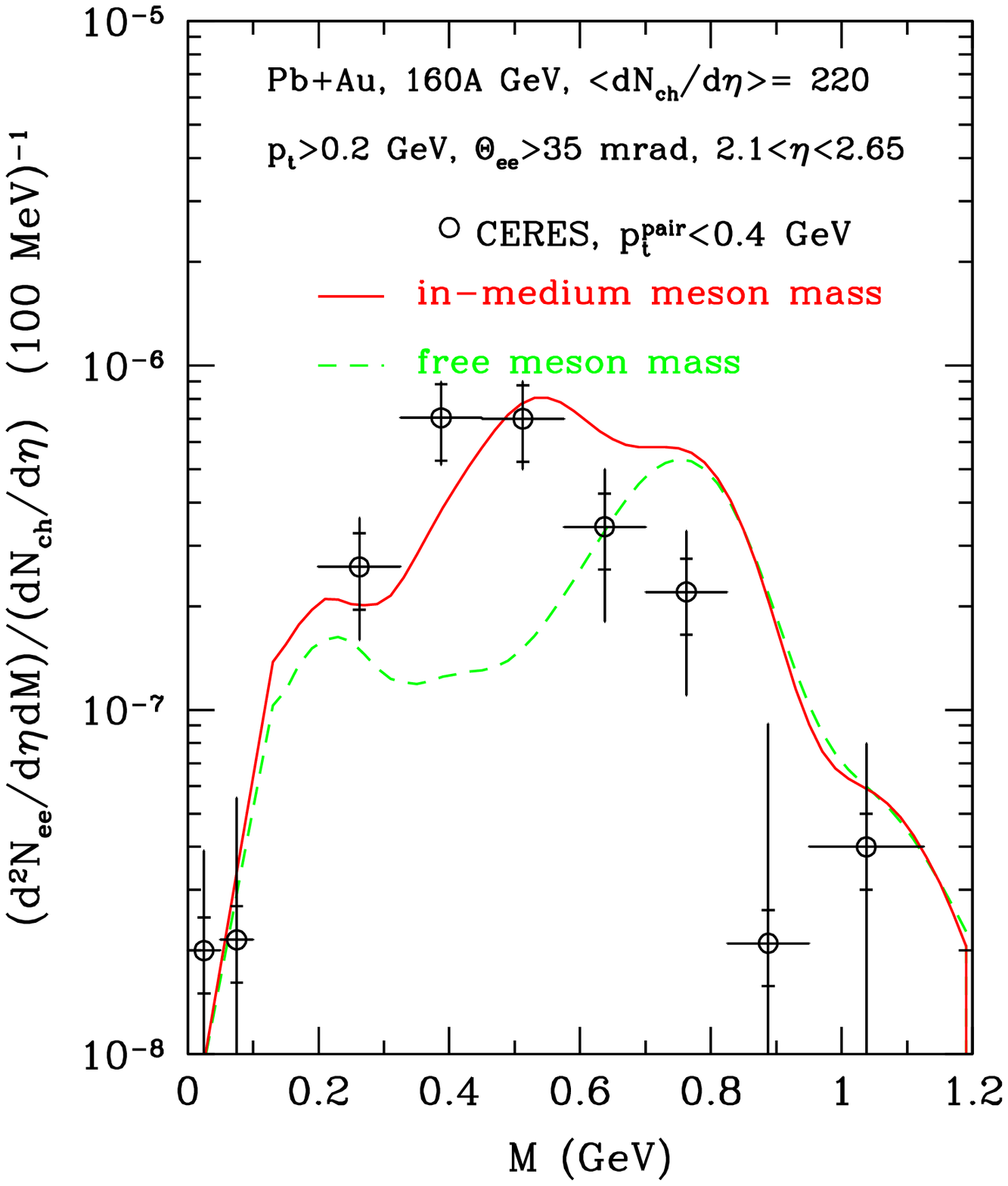,width=6.7cm,height=6cm}}}  
\caption{Dilepton mass spectra with $p_t^{pair}<$ 0.4 GeV in
Pb+Au collisions at 158 GeV/u compared with: (a) fireball calculations 
by R. Rapp (left panel) using the R/W spectral function (full curve), 
B/R scaling (dashed curve) and without medium effects (dotted curve);
Dalitz decay contributions are not included here; 
(b) transport calculations by G.Q. Li (right panel) using 
B/R scaling (full curve) and without medium effects (dashed curve). }
\end{figure}
We shall concentrate on the energy region around $M_{ee}=m_\rho/2$.
This is sensibly above the $\pi^0$ background peak, and yet low
enough to make it impossible for theories without dropping
$\rho$-mass to achieve the observed enhancement. 
It should be noted that the Rapp curve is a factor $\sim$~2 below the 
experimental points.
Note also that the curve with {\it no} medium effects in the $\rho$-mass lies
an order-of-magnitude below the data.

The near agreement of the current Rapp/Wambach calculations with the 
empirical $p_\perp<400$~MeV/c-spectrum shows in terms of 
eq.~(\ref{scaling}) that their ${\vec q}^2$ term is rather small 
compared with the effective mass term (corresponding to the 
second term on the RHS of eq.~(\ref{scaling})). So, 
directly from the inspection of this curve we can see that they 
have a rather small $m_\rho^*$ at the relevant densities. 
In the right panel of fig.~1 are shown results calculated by G.Q. Li 
using the Li-Ko-Brown theory~\cite{LKB}. Here, the Dalitz pairs
are included. The free meson mass curve is shown to be still
a large factor, $\sim$~5 too low. The curve with in-medium meson
masses is similar in shape to that of Rapp, but a factor $\simeq$~2
higher than his curve. In fig.~2 the $p_\perp$ distribution from
Li-Ko-Brown theory is shown to generally provide a good fit.

\begin{figure} 
\centerline{\epsfig{file=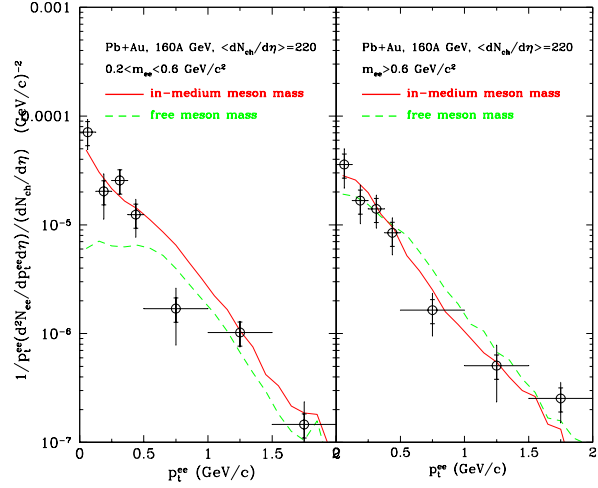,height=4in,angle=-90}}
\caption{Dilepton $p_t$ spectra in two invariant mass bins in 
Pb+Au collisions at 158 AGeV compared with the transport 
calculations by G.Q. Li using B/R scaling (full curves) and 
without medium effects (dashed curves).} 
\end{figure}

In the following we show a crucial ingredient in the Rapp/Wambach
dilepton enhancement is a dynamical lowering of the $\rho$-meson mass,
accomplished by $S$-wave interactions coupling to
$N^*$-hole states.

Rapp et al.~\cite{RUBW} in preparation for including scalar isobar 
excitations in the dilepton analysis, have calculated photoabsorption
cross sections on protons and nuclei. The (with respect to dilepton
production) most important excitation is that of the $N^*(1520)$, the 
other $S$-wave excitations being smaller. In the present R/W theory
the latter may make up an $\sim$~30\% correction to that from the 
$N^*(1520)$. However, we believe that only the latter should be 
multiplied by the full baryon -- rather than nucleon -- density,
since the dipole excitations can be built upon excited nucleon states.
Thus, the importance of the $N^*(1520)$ in their theory, compared with 
the non-collective excitations, should be greater.  
Using a phenomenological 
Lagrangian of the form~\cite{PPLLM,RUBW} 
\beq
{\cal L}_{\rho N^*N}^{s-wave}  =  \frac{f_{\rho N^*N}}{m_\rho} \
\Psi^\dagger_{N^*} \ (q_0 \ {\vec s}\cdot \vec{\rho}_a -
\rho^0_a \ {\vec s}\cdot {\vec q}) \ t_a \ \Psi_N \ + \ h.c.\label{phen} \
\eeq    
they determine 
\beq
f_{\rho N^*N}^2/4\pi = 5.5 
\eeq
from fits to the photoexcitation data. In addition to the 
$\Gamma_0\simeq$~120~MeV
natural width of the $N^*(1520)$ they include an additional medium 
dependent width of 250~MeV for nuclear matter density. 

Rapp et al. diagonalize the interactions leading to several
$N^*$ resonances. Here we make a 
schematic calculation of the excitation 
of the $N^*(1520)$, which gives the dominant contribution. 
Consider the two-level toy model of the $N^*(1520)$ and the $\rho$.
The two branches in the $\rho$ spectral function can
be located by solving the (real part of) the  $\rho$-meson dispersion 
relation (at $\vec{q}=0$),  
\beq
q_0^2=m_\rho^2+{\rm Re}\Sigma_{\rho N^*N}(q_0) \ , 
\label{disp} 
\eeq
self-consistently. Including also the backward-going graph 
the $N^*(1520)N^{-1}$ excitation contributes to the self-energy  
at nuclear matter density $n_0$
\beq
\Sigma_{\rho N^*N}(q_0)= f_{\rho N^*N}^2 \ \frac{8}{3} \ 
\frac{q_0^2}{(m_\rho)^2} \ \frac{n_0}{4} \  
 \left( \frac{(\Delta E)^2}{(q_0+i\Gamma_{tot}/2)^2-(\Delta E)^2} 
\right)
\label{sigma} 
\eeq
where $\Delta E\approx 1520-940=580$ MeV and 
$\Gamma_{tot}=\Gamma_0+\Gamma_{med}$ is the total width of the $N^*(1520)$. 
When neglecting any in-medium corrections to the width of the $N^*(1520)$ 
we find two solutions for eq.~(\ref{disp}), located at  
\beq
q_0^-\simeq540{\rm MeV} \quad , \quad q_0^+\simeq 895{\rm MeV} \ .   
\label{solutions} 
\eeq
Solving eq.~(\ref{disp}) is in fact equivalent to determining the 
zeros in the real part of the $\rho$-meson propagator,
\beq
D_\rho(q_0,q)=1/[q_0^2-q^2-m_\rho^2-\Sigma_{\rho N^*N}(q_0,q)+im_\rho
\Gamma_\rho] 
\eeq
at vanishing 3-momentum (cp. long-dashed line in the lower left panel of
fig.~3). However, it is more instructive to  
examine the imaginary part of $D_\rho$ which is directly related
to the $\rho$-meson spectral function $A_\rho=-2{\rm Im}~D_\rho$.  
For $\Gamma_{med}$=0 we clearly see the two states corresponding 
to the two solutions quoted above (long-dashed line in the upper
left panel of fig.~3), with the low-lying peak actually situated at 
$q_0=$500~MeV.  Although for finite $\Gamma_{med}$ the low-lying solution 
disappears (short-dashed line in the lower left panel of
fig.~4), the $\rho$-spectral function  still exhibits 
appreciable strength around $q_0\simeq$550~MeV, {\it i.e.}~the low-lying
state simply broadens thereby losing some of its collectivity 
(short-dashed curve in the upper left panel of fig.~3). 
However, only a fraction of the $\rho$-meson strength 
resides in this lower state ($Z_-\simeq0.2$). 
\begin{figure}
{\makebox{\epsfig{file=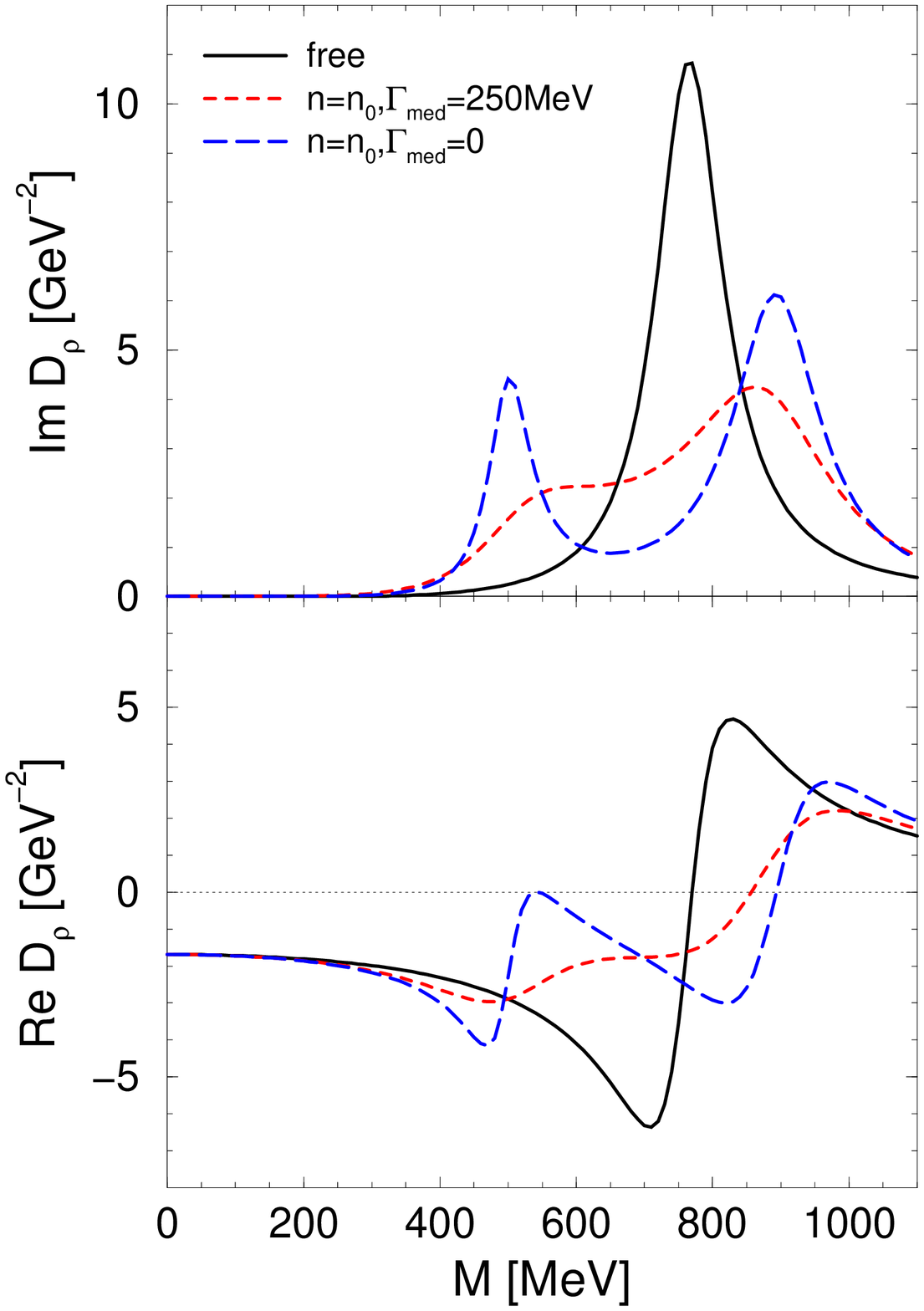,width=6.25cm,height=10cm}}}
{\makebox{\epsfig{file=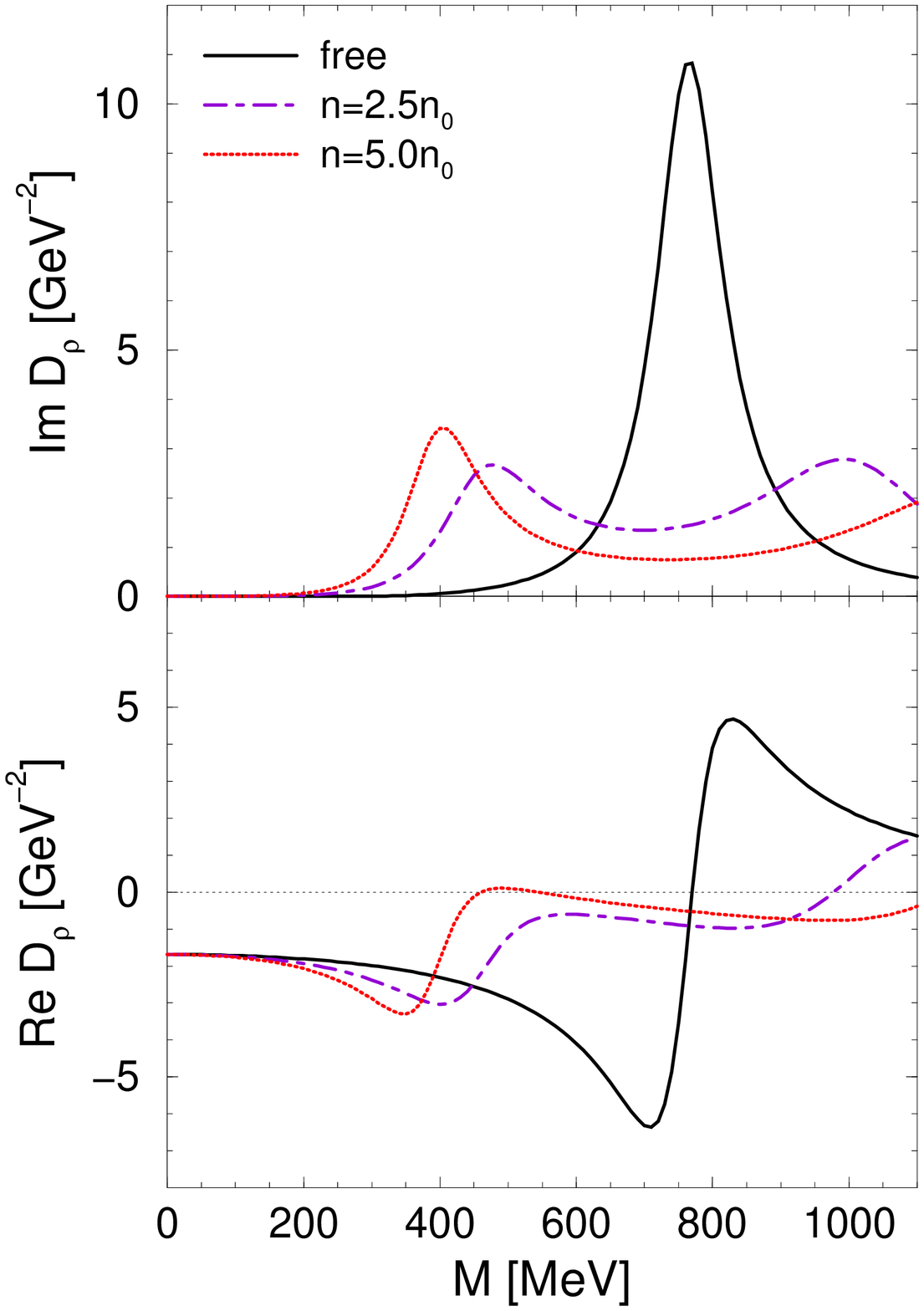,width=6.25cm,height=10cm}}}
\caption{Real part (lower panels) and imaginary part (=spectral 
function, upper panels) of the $\rho$-meson propagator when 
including $N^*(1520)N^{-1}$ excitations according to eq.~(\ref{sigma});
left panel: at $n$=0 (full curves) and at normal nuclear matter 
density ($n$=$n_0$) for two
different values of the in-medium $N^*(1520)$ width, namely
$\Gamma_{med}$=0 (long-dashed curves) and $\Gamma_{med}$=250~MeV
(short-dashed curves); right panel: development of the collective state 
at higher nuclear densities, {\it i.e.}~$n$=2.5$n_0$ (dashed-dotted curve) 
and $n$=5$n_0$ (dotted curve), both for $\Gamma_{med}$=250~MeV.} 
\end{figure}
This constitutes a significant difference from B/R scaling
in which the scalar meson field brings the $m_\rho^*$ down to
$m_\rho/2$ for $n\simeq n_0$, but leaves the full pole strength.
Although R/W lose dileptons, as compared with B/R scaling, 
they include many other channels such as further scalar excitations
etc.. This leads to a final discrepancy of a factor $\sim$2 
as can be roughly inferred from the peak in the $p_\perp<400$~MeV/c 
$M$-spectrum  of Rapp in fig.~1, which is about half of that of the 
calculation using B/R scaling. However, the R/W peak will 
be substantially increased if the 
full baryon density is used in calculating effects of the $N^*(1520)$, 
and the relative role of the other effects will be smaller 
(note that their use of nucleon -- rather than baryon --
density is probably more correct for the non-collective $S$-wave
resonances).

\section{Collective modes more generally}

Nuclear giant resonances are often produced by operating on the 
``vacuum" ({\it i.e.}, 
nucleus in its ground state) by the appropriate operator. 
Thus, the giant dipole state ($GDR$) is typically constructed as 
\beq
\Psi(GDR)\simeq D \Psi_0 \ , 
\eeq
where $D=N\sum z_i\tau_i$ with $N$ the normalization constant. 
$\Psi(GDR)$ is a coherent sum over particle-hole excitations. 

In a similar way, the collective $N^*$-hole state with quantum
numbers of the $\rho$-meson is
\beq
\Psi_\rho=\rho|0\rangle \simeq [N^*(1520) N^{-1}]^{1^-} \ , 
\label{Psirho}
\eeq
since each $N^*$-hole excitation has the same energy. In fact 
$N^*(1520)$ decays $\sim$~80\% of the time by $\pi$ or $2\pi$
emission, and the dispersion from this decay lowers its 
energy by 190~MeV, cutting down the decay into the $\rho$-channel
by the factor 
\beq
F=\frac{(\Gamma_\rho/2)^2}{(190{\rm MeV})^2+(\Gamma_\rho/2)^2} =
\frac{1}{7.4}  
\eeq
from what the decay to the $\rho$ would be were the $N^*(1520) N^{-1}$  
at the $\rho$-pole. There is an additional factor of 3/8 for the 
proportion of the direct product $N^*(1520) N^{-1}$ which is 
coupled to $J^P=1^-$. Thus, the internal structure of 
$[N^*(1520) N^{-1}]^{1^-}$ is estimated to have fraction 
\beq
f>\frac{7.4\times8/3\times0.2}{0.8+7.4\times8/3\times0.2} =0.83\label{f} \ , 
\eeq
where the 0.8 in the denominator is the 
observed fraction of $N^*(1520)$ decay into 
$N\pi$ or $N\pi\pi$. The $f$ is greater in (\ref{f}) because lowering the
energy of the $N^*(1520) N^{-1}$ state from that of the $\rho$ will
decrease the pion decay. Our conclusion is that the 
$[N^*(1520) N^{-1}]^{1^-}$ is about as close to $\rho|0\rangle$ 
as the $D\Psi_0$ is to the $\Psi(GDR)$ eigenstate. The difference is 
that the elementary particle corresponding to the $GDR$ is the photon,
whose zero mass is protected by gauge invariance whereas here the nonabelian
gauge particle ({\it i.e.}, $\rho$) is massive (which may be considered as due
to Higgs mechanism a la spontaneously broken hidden gauge symmetry). 

One might ask why the $N^*(1520)N^{-1}$ comes at an energy 190~MeV below
that of a free $\rho$. It is easy to see that the open $\pi$ and $2\pi$
decay channels have dispersion corrections which move the 
energy of the $N^*(1520)N^{-1}$ down. Taking these to come from
loop corrections involving the off-shell decay of the $\rho$, one
sees that there is a certain self-stability, in that if the 
$N^*(1520)$ drops too far in energy, the pion decay will be 
cut off. The detailed evaluation of the dispersion correction 
remains to be done, however.  

Consequently, in the case of the $\rho$-meson, the elementary $\rho$    
and the nuclear collective state 
$\rho|0\rangle=\sum [N^*(1520) N^{-1}]^{1^-}$  will 
interact strongly, the symmetrical contribution moving to
energy $\sim m_\rho/2$. The baryons, neutron and proton, 
have isospin degeneracy 2, so one can build up two nuclear 
vector mesons. The coupling between
elementary $\rho$ and $N^*$-nucleon-hole $\rho$ should be universal. 

\section{Broadening} 

One of the chief differences between R/W theory and that 
following from B/R scaling is that the $\rho$-meson is very broad
in the former whereas Li, Ko and 
Brown~\cite{LKB} worked in mean field approximation. Ko~\cite{Ko}  
has added effects of collision broadening to the latter theory. 
In lowest order the effects such as $\pi$-$\rho$ scattering give 
a large width in the denominator of the $\rho$-propagator because
the lifetime of the wave packet representing the $\rho$ is terminated
by such a collision. The short lifetime is accompanied by a large width.
Thus, the contribution from lowest order is substantially reduced, due
to the large $\Gamma_{med}$. However, in next order the dilepton
production from the $\pi$-$\rho$ interaction adds to the numerator, 
compensating for the decrease due to the introduction of the large
$\Gamma_{med}$ in lowest order. To the extent that the additional terms
added to the numerator compensate for the width in the denominator, the
broadening therefore does not change the situation.

\section{Conclusion} 
We have shown that a large contribution to the low-mass dilepton spectrum
in the R/W scenario comes from the elementary $\rho$ mixing with 
the $[N^*(1520) N^{-1}]^{1^-}$  state. This produces a strong 
elementary $\rho$ component at an energy $\sim m_\rho/2$ for nuclear 
matter density $n_0$. 

The $[N^*(1520) N^{-1}]^{1^-}$ has as main component the giant resonance 
$\rho|0\rangle$. In this sense it is analogous to the particle-hole
combination of the giant dipole resonance in nuclear theory, 
$\Psi(GDR)=D|0\rangle$, where $D$ is the dipole operator and 
$|0\rangle$ the ground state of the nucleus. Thus, the
$[N^*(1520) N^{-1}]^{1^-}$ is chiefly a coherent state of $N^*$-hole
combinations, summed over all nucleon-holes, with the quantum
numbers of the $\rho$-meson. Many other small components of the 
in-medium $\rho$ also contribute in R/W.

An important ingredient in the R/W
calculations leading to low-mass dilepton enhancement is the low-mass
component of the $\rho$ at around 500~MeV
for nuclear matter density $n_0$ (this component has, however, 
a strength of only $\sim$~20\% of the bare $\rho$,  
although many other components are mixed through the spectra). 

Moving to higher densities/temperatures we claim that the full baryon 
density $n_B$ should be used in calculating $\Sigma_{\rho N^*N}$, 
because excited baryons 
will also have a dipole state built on them. This may not have quite 
the same regularity as the dipole state built on the nucleons. Also
finite temperature effects may enter. Thus we expect to somewhat 
overestimate the rate at which the symmetric coherent state 
moves down in energy. From $m_\rho^*=$550~MeV at nuclear matter 
density this state moves down further with density. 
At baryon densities of around 5$n_0$, which roughly correspond
to initial conditions in central 160GeV/u Pb+Au, it has reached 
about 400~MeV (see right panel of fig.~3). This seems not enough
to provide a direct link to B/R scaling, and can be traced
back to the fact that the preservance of gauge invariance at the 
hadronic level requires derivative coupling of the $\rho$ to the 
nucleons (resulting in the $q_0^2$-factor in eq.~(\ref{sigma}),
which slows down the rate of decrease in mass appreciably). 
This feature is not present in the B/R scenario (in which the factor 
$q_0^2/m_\rho^2$ in eq.~(\ref{sigma}) would be replaced by unity), 
which, however, is
chiefly designed to work at the constituent quark level.  
Working with constituent quark degrees of freedom, Brown, Buballa 
and Rho found the phase transition to occur around 
$n_c=2.75\rho_0$~\cite{BBR}, {\it i.e.}~the $\rho$-meson becoming
massless at this point.  
To the extent that this results, the changeover from 
hadronic to constituent quark language should be made rather early 
for densities not very far above $n_0$. Below $n_0$ the hadronic 
language should certainly be used. In this sense, the question 
may be no longer which is the 'correct' scenario, but at precisely
what densities does the hadronic description break down. 
This may in fact be studied experimentally in terms of systematic
centrality dependencies in the dilepton yield in a quantitative way. 
The explicit demonstration of quark-hadron duality in this context,   
which has to show up in some 
intermediate density regime, remains a theoretical challenge.  

Whereas the first author, in writing this paper, has extolled the
virtues of B/R scaling, it should be recognized that R/W have 
provided us with close connections with many observable phenomena
in the hadronic world. We have learned that form factors of
$P$-wave couplings must be drastically softened, firstly in order
to reproduce known two-body collision data~\cite{Fr97}, but 
generally so as to make the ${\vec q}^2$ term in eq.~(\ref{scaling}) 
small, essentially unimportant. The analysis~\cite{RUBW} found
the $f_{\rho N^*N}$ to the $N^*(1520)$ resonance to be very close to that
we find from universal vector meson coupling. In general, it has
been useful to give new and improved understanding of the many
empirical processes which figure in the R/W description. It is
highly satisfying that R/W and B/R end up in qualitative 
agreement\footnote{Quantitatively there are one or two critical points 
of difference which can now be focused on.}. It is 
remarkable that what started out to be a highly complicated 
dynamical mechanism in R/W  might turn out to have an extremely 
simple picture in terms of weakly interacting quasiparticles. This then
provides evidence for correspondence between the hadronic picture and
the partonic (quasiparticle) picture in the high density regime 
comparable in quality to what one gets at nuclear matter density 
when probed by  very low energy electroweak probes, such as static
moments of heavy nuclei, nuclear axial-charge transitions and 
nuclear matter properties~\cite{SBMR}. One plausible way of making 
the crossover from the low to high density regime would be
to shift from {\it nuclear} Fermi liquid structure to {\it quasiquark} Fermi 
liquid, with the collective state in
the $\rho$-meson channel constructed above being an analog to
 zero-sound excitation
on top of the Fermi liquid ground state defined at a given density.  

Furthermore, B/R suggested their scaling as only approximate. It is 
essentially an effective theory treated at tree order. Some of the effects
taken into account in R/W may not be included in B/R and vice versa.
Considering  other mesons such as the $\omega(783)$ we
should be able to calculate differences in scaling, that is, corrections
to the tree order results
(note, {\it e.g.}, that Lee~\cite{Lee} has a much greater 
momentum dependence in $m_\omega^*$ than in $m_\rho^*$). We have
also been able to identify many other baryon resonances in the 
particle data tables which are likely to be dipole excitations
on lower-mass resonances.  
We believe that what we have learned here will be very useful in
calculating the deviation from B/R scaling.

\subsection*{Acknowledgments} 

We would like to thank U. Mosel for pointing out
the importance of the $N^*(1520)$ in the dynamical mechanism.
One of us (RR) acknowledges partial support from the 
A.-v.-Humboldt-foundation as a Feodor-Lynen-fellow. This work is 
supported in part by the U.S. Department of Energy under grant
No. DE-FG02-88ER-40388.


\begin{thebibliography}{9}
\bibitem{sau} {G. Agakichiev {\it et al.}, CERES collaboration},
Phys. Rev. Lett. {\bf 75} (1995) 1272.
\bibitem{RCW} R. Rapp, G. Chanfray and J. Wambach, Nucl. Phys. {\bf A617}
(1997) 472; \\
R. Rapp, Proceedings of the 33rd Recontres de Moriond on ``QCD and High Energy
Hadronic Interactions'', Les Arcs (France), March  21-28, 1998, 
to be published in Edition Frontiers, ed. J. Tran Thanh Van, and 
nucl-th/9804065.  
\bibitem{BRscaling} G.E. Brown and M. Rho, Phys. Rev. Lett. {\bf 66} (1991) 
2720. 
\bibitem{LKB} G.Q. Li, C.M. Ko and G.E. Brown, Nucl. Phys. {\bf A606}
(1996) 568.
\bibitem{HaLe} T. Hatsuda and S.H. Lee, Phys. Rev. {\bf C46} (1992) R34.
\bibitem{JiLe} X. Jin and D.B. Leinweber, Phys. Rev. {\bf C52} (1995) 3344.
\bibitem{BBR} G.E. Brown, M. Buballa and M. Rho, Nucl. Phys. 
{\bf A609} (1996) 519.
\bibitem{SBMR} Chaejun Song, G.E. Brown, D.-P. Min and M. Rho, Phys. Rev.
{\bf C56} (1997) 2244; Chaejun Song, D.-P. Min and M. Rho, Phys. Lett. 
{\bf B424} (1998) 226.  
\bibitem{qhduality} See, e.g., B. Grinstein and R.F. Lebed, ``Explicit 
quark-hadron duality in heavy-light meson weak decays in the 't~Hooft model,''
hep-ph/9708396. 
\bibitem{FR96} B. Friman and M. Rho, Nucl. Phys. {\bf A606} (1996) 303;
B. Friman, M. Rho and C. Song, ``Chiral Lagrangians and Landau Fermi liquid
theory for dense matter'', to appear; M. Rho, hep-ph/9711466
\bibitem{wilczek} M. Alford, K. Rajogopal and F. Wilczek, hep-ph/9711395; 
hep-ph/9804403; \\
R. Rapp, T. Sch\"afer, E. Shuryak and M. Velkovsky, hep-ph/9711396.  
\bibitem{Voigt} C. Voigt, PhD thesis, University of Heidelberg, Heidelberg
1998. 
\bibitem{FrPi} B. Friman and H.J. Pirner, Nucl. Phys. {\bf A617} (1997)
496.
\bibitem{Fr97} B. Friman, in Proceedings of the APCTP Workshop
on 'Hadron Properties in Medium', Seoul, Korea, Oct. 27-31, 1997, 
to be published.
\bibitem{Lee} S.H. Lee, Phys. Rev. {\bf C57} (1998) 927. 
\bibitem{RUBW} R. Rapp, M. Urban, M. Buballa and J. Wambach, 
Phys. Lett. {\bf B417} (1998) 1.
\bibitem{PPLLM} W. Peters, M. Post, H. Lenske, S. Leupold and U. Mosel,
Nucl. Phys. {\bf A632} (1998) 109. 
\bibitem{Ko} C.M. Ko, private communication and in preparation. 
\bibitem{BLL} G.E. Brown, C.-H. Lee and G.Q. Li, in Proceedings of the 
APCTP Workshop on 'Hadron Properties in Medium', Seoul, Korea, Oct. 27-31, 
1997, to be published. 
\bibitem{BR96} G.E. Brown and M. Rho, Phys. Rep. {\bf 269} (1996) 333. 
\bibitem{Brown} G.E. Brown, Chap. V.3 in {\it Unified Theory of Nuclear Models
and Forces}\ ( North Holland Publ.-Company, Amsterdam-London 1971).  
\end{thebibliography}
\end{document}